\documentclass[11pt]{amsart}
\usepackage{hyperref}
\font\gr=cmr10 scaled 1100
\def \dd{{\rm d}}
\def \DD{{\rm D}}
\begin{document}

\title[David Hilbert and the origin of the ``Schwarzschild solution'']
{David Hilbert and the origin of the ``Schwarzschild solution''}
\author{Salvatore Antoci}%
\address{Dipartimento di Fisica ``A. Volta'' and INFM, Pavia, Italia}
\email{Antoci@matsci.unipv.it}

\begin{abstract}
The very early dismissal of Schwarzschild's original solution
and manifold, and the rise, under Schwarzschild's name, of
the inequivalent solution and manifold found instead by Hilbert,
are scrutinised and commented upon, in the light of the
subsequent occurrences.

    It is reminded that Hilbert's manifold suffers from two defects,
that are absent in Schwarzschild's manifold. It does not admit a
consistent drawing of the arrow of time, and it allows for an
invariant, local, intrinsic singularity in its interior. The
former defect is remedied by the change of topology of the
extensions proposed by Synge, Kruskal and Szekeres. The latter
persists unaffected in the extensions, since it is of local
character.
\end{abstract}
\maketitle
\section{Introduction}
There is, indisputably, no issue in Einstein's theory of general
relativity that has been so accurately scrutinized, and by so many
relativists in so many decades, like the ``Schwarzschild solution''.
Innumerable research articles have been devoted to its study, and still are
at the present day. Any textbook of relativity, either introductory
or advanced, cannot help dedicating one chapter or more to the
derivation of this paradigmatic solution of Einstein's field equations
for the vacuum, followed by a discussion of the obtained result and of the
theoretical predictions stemming from it. In the books published
after 1970 (with some notable exceptions) one more chapter is then
devoted to the task of removing, through the Kruskal maximal extension, the
singularity that the metric components in the famous ``Schwarz\-schild''
expression for the interval
\begin{equation}\label{1}
\dd s^2=\left(1-\frac{2m}{r}\right)\dd t^2
-\left(1-\frac{2m}{r}\right)^{-1}{\dd r^2}
-r^2(\dd \vartheta^2+\sin^2\vartheta \dd \phi^2),
\end{equation}
where
\begin{equation}\label{2}
0<r<\infty,
\end{equation}
exhibit, in ``Schwarzschild'' coordinates, at $r=2m$. The reader is always
ensured that this is a spurious singularity, devoid of local physical
meaning, due only to the bad choice of the coordinate system done by
Schwarzschild.

    It is therefore a bit surprising to learn, through the direct reading
of the original ``Massenpunkt'' paper \cite{Schwarzschild16a}, that
Karl Schwarzschild never wrote a solution given by equations
(\ref{1}) and (\ref{2}), nor a solution whose manifold was in one
to one correspondence with the latter. Even worse: it turns out
that, due to his method of solution, he had the possibility to
write a manifold in one to one correspondence with the manifold
described by (\ref{1}) and (\ref{2}), but deliberately refused
to do so.

In fact, after the Minkowskian boundary
conditions at the spatial infinity have been satisfied, Schwarzschild's original
solution appears to contain still two constants of integration, instead
of the single one that appears in (\ref{1}) and (\ref{2}). One of
the constants has to do with the active gravitational mass, and
Schwarzschild chose it by appealing to Newton; the second
one determines the position of the inner border of the manifold.
Schwarzschild therefore needed an additional postulate in order to fix this
second constant. By appealing to the requirement of continuity for
the components of the metric in the range between the inner border
and the spatial infinity, Schwarzschild chose his second constant
in such a way as to position the singularity that brings his name
just on the inner border of the manifold.

This singular outcome of the perusal of Schwarzschild's original
paper \cite{Schwarzschild16a} will not be expounded here any further,
because it has already been scrutinized in the Note
\cite{Note2003} that accompanies a recent English translation
\cite{Schwarzschild2003} of the ``Massenpunkt'' paper. One has
rather answering here the ensuing questions: how did it happen that
the manifold described by (\ref{1}) {\it cum} (\ref{2}) was called
``Schwarzschild solution'', and why and when the original solution with two
constants of integration, hence with the need for an additional
postulate, was forgotten\footnote{The memory of Schwarzschild's
original solution was rekindled at the end of the last century
by the works \cite{Abrams79,Abrams89,Pekeris82}
of L. S. Abrams and  C. L. Pekeris.}?

\section{Frank's review of the ``Massenpunkt'' paper}
It is remarkable that the substitution of the ``Schwarzschild''
solution (\ref{1},\ref{2}) for the original one \cite{Schwarzschild16a}
war a very early occurrence, certainly easied by the premature
death of Karl Schwarzschild. The seeds of oblivion were already cast
in the review with which Philipp Frank presented \cite{Frank1916}
Schwarzschild's ``Massenpunkt'' paper to the community of the mathematicians.
An English translation of the review is reported in Appendix A .
The interested reader is invited to compare that necessarily
concise account with the original paper \cite{Schwarzschild16a}.
In this way one can appreciate that Frank's review faithfully
extracts several relevant points of Schwarzschild's achievement
by accurately following the letter of the text. For the sake of
conciseness, however, two facts were completely left in the
shadow. Their omission might have appeared marginal to Frank at the time the
review was written, but it  became crucial soon afterwards, when the
rederivations of the Schwarzschild solution by Droste, Hilbert and Weyl
appeared \cite{Droste17, Hilbert17,Weyl17} in print. And today, if
one reads Frank's account without having previously perused
Schwarzschild's paper \cite{Schwarzschild16a}, one by no
means understands the rationale of Schwarzschild's procedure, and
why the manifold found by him happens to be inequivalent to the one found
in particular by David Hilbert.

By reading the review, one agrees of course with the initial choice of the
interval, depending on three functions $F(r)$, $G(r)$, $H(r)$, but is soon
led to wonder why Schwarzschild did abandon the polar
coordinates $r$, $\vartheta$, $\varphi$, that he had just introduced,
and felt the need to go over to new spatial coordinates $x_1$, $x_2$,
$x_3$, defined by the transformation:
$$
x_1=\frac{r^3}{3}, \quad x_2=- \cos \vartheta,
\quad x_3=\varphi.
$$
One then wonders how Schwarzschild could determine his
three new unknown functions $f_1$, $f_2=f_3$, $f_4$ from Einstein's
field equations without imposing one coordinate condition, that
is not mentioned in Frank's account.
Only by looking at the reviewed paper does one gather why
Schwarzschild did work in that way. One discovers that
he did not solve the field equations of the final version
\cite{Einstein15c} of Einstein's theory, but the equations of the
previous version, that Einstein had submitted \cite{Einstein15a}
to the Prussian Academy of Sciences on November 11th, 1915. Those
equations provided for the vacuum the same solutions as the final ones,
but limited the covariance to unimodular coordinate
transformations. They read:
\begin{equation}\label{3}
\sum_\alpha\frac{\partial \Gamma^\alpha_{\mu\nu}}
{\partial x_\alpha}
+\sum_{\alpha\beta}~\Gamma^\alpha_{\mu\beta
}\Gamma^\beta_{\nu\alpha}=0,
\end{equation}
and
\begin{equation}\label{4}
\vert g_{\mu\nu}\vert=-1.
\end{equation}
This very fact explains why Frank did not mention any coordinate
condition in his review: Schwarzschild did not need one, for
equation (\ref{4}), meant by Einstein as a further field equation,
played that r\^ole. This circumstance explains too the otherwise
mysterious adoption of the coordinates  $x_1$, $x_2$, $x_3$, named
by Schwarzschild \cite{Schwarzschild16a}
``polar coordinates with determinant 1".

While the mentioned omission only hampers a proper understanding
of Schwarz\-schild's procedure, the second omission had more far
reaching consequences. It concerns the number of nontrivial constants of
integration that occur in Schwarzschild's integration of
(\ref{3}) and (\ref{4}). They are three, and were pointedly
labeled as $\alpha$, $\rho$ and $\lambda$ integration constants
by the careful Schwarzschild, but only $\alpha$ appears in the
final result reported by Frank. In the review, no word is spent
about the existence of two more constants of integration, and
about the way kept in fixing them.
The omission is nearly irrelevant for $\lambda$, since
Schwarzschild just set $\lambda=1$ in order to fulfill both the
Minkowskian boundary condition at infinity and the requirement that,
for vanishing mass, the Minkowski manifold be retrieved.
It is however crucial for $\rho$. The functions $f_i$,
as they read before fixing $\rho$, are given in equations (10)-(12)
of the ``Massenpunkt'' paper \cite{Schwarzschild16a}. They are:
\begin{eqnarray}\nonumber
f_1=\frac{(3x_1+\rho)^{-4/3}}
{1-\alpha(3x_1+\rho)^{-1/3}},\\\label{5}
f_2=f_3=(3x_1+\rho)^{2/3},\\\nonumber
f_4=1-\alpha(3x_1+\rho)^{-1/3}.
\end{eqnarray}
Schwarzschild notes that (\ref{5}) satisfy all the conditions
previously postulated for the solution, except for the  condition of
continuity, because $f_1$ is discontinuous when
\begin{equation}\label{6}
1=\alpha(3x_1+\rho)^{-1/3}, \ \ \text{\it i.e.} \ \ 3x_1=\alpha^3-\rho.
\end{equation}
In order that this discontinuity coincides with the origin of $x_1$,
namely, with the inner border of the manifold considered by him,
Schwarzschild chose
\begin{equation}\label{7}
\rho=\alpha^3.
\end{equation}
This is the reason why only the integration constant $\alpha$ survives
in the final result reported by Frank. His review, however, by no means
tells the reader that a problem, which had to become of fundamental importance
for the future generations of relativists, had been seen by Schwarzschild,
and deliberately solved in a certain way.

\section{Hilbert's rederivation of the Schwarzschild solution}
Frank's review of 1916 by its omissions certainly did not help
in providing mathematicians and physicists alike with a clear idea
both of the major issue that
Schwarzschild had confronted when first solving Einstein's
vacuum equations for the spherically symmetric, static case, and of
the way out chosen by him. It was however Hilbert, with his
revisitation \cite{Hilbert17} of the static, spherically
symmetric problem, published in 1917, that definitely imposed the ostracism
on the original Schwarzschild solution. He did so by attaching the name of
Schwarzschild to the metric and the manifold defined by (\ref{1})
and (\ref{2}), that were instead the outcome of his own work, while
dismissing in a footnote as ``not advisable'' the inequivalent,
pondered choice of the manifold done by Schwarzschild. To document this
occurrence, an English translation of the excerpt from Hilbert's
paper \cite{Hilbert17} that deals with the rederivation of the
spherically symmetric, static solution is reported
in Appendix B; the above mentioned footnote is just at the
end of the excerpt.

    It must be aknowledged that, in this occasion, destiny exerted
some of the irony of which it is capable with the great David Hilbert.
In fact, as rightly noted \cite{Abrams89} by Leonard Abrams, in the very
paper by which he condemned Schwarzschild's deliberately chosen
manifold to undeserved oblivion, Hilbert committed an error.
A crucial constant of integration, that played in Hilbert's procedure
just the r\^ole kept by $\rho$ in Schwarzschild's calculation, was unknowingly
allotted by him an arbitrary value, thereby fixing by pure chance
the manifold in the ``Schwarzschild'' form (\ref{1}) plus (\ref{2}).
Hilbert's error was no doubt influential in rooting in many a relativist
the wrong conviction that the manifold defined by (\ref{1}) and (\ref{2})
is a necessary outcome of the field equations of general relativity.
Indeed, it corresponds just to one particular way of choosing the position
of the inner border, that could have been adopted by Schwarzschild too, had he
renounced his injunction of continuity for $f_1$, and chosen
$\rho=0$ instead of $\rho=\alpha^3$.

Let us consider Hilbert's derivation in some detail.
In the footsteps of Einstein and Schwarzschild, he first postulated
the conditions that the line element must obey, when written with
respect to ``Cartesian'' coordinates,
in order to describe a spherically symmetric, static
manifold. Then he went over to polar coordinates and wrote the
line element (\ref{42}), where $F(r)$, $G(r)$ and $H(r)$ are three
unknown functions. Due to the general covariance of the final
field equations \cite{Einstein15c,Hilbert15} of the theory, that he himself had
contributed to establish, in order to write a solution
exempt from arbitrary functions, one must impose  on the line element (\ref{42})
one coordinate condition, that reduces the number of the unknown functions to two.
Hilbert decided to fix $G(r)$ by introducing a new radial coordinate
$r^{\ast}$, such that
$$
r^{\ast}=\sqrt{G(r)}.
$$
He then dropped the asterisk, thereby
writing the line element (\ref{43}), that contains only two unknown
functions, $M(r)$ and $W(r)$, of the ``new'' $r$, and constitutes
the canonical starting point for all the textbook derivations
of the ``Schwarz\-schild solution''.
This is quite legitimate. What is not legitimate, although first done by
Hilbert and subsequently handed down to the posterity,
is to assume without justification that the range of the ``new'' $r$ is still
$0<r<\infty$, as it was for the ``old'' $r$, because this is tantamount to setting
$\sqrt{G(0)}=0$, an arbitrary choice \cite{Abrams89}, equivalent to
setting $\rho=0$ in Schwarzschild's result, reported in equation (\ref{5}).
\section{Forethoughts and afterthoughts}
It might be asserted that Hilbert's error, when compared to
Schwarz\-schild's meditated option for continuity, was a sort of
{\it felix culpa}, because it was, perhaps by prophetic inspiration,
fully in line with the subsequent understanding gained when
the intrinsic viewpoint of differential geometry was correctly applied
to general relativity. Through this improved understanding
the discontinuity of $f_1$ occurring when $\ 3x_1=\alpha^3-\rho$,
that so much bothered Schwarzschild as to induce him to decide its
relegation to the inner border of the manifold by setting $\rho=\alpha^3$,
revealed itself to be a mere coordinate effect. On the contrary, the
singularity occurring at $r=0$ in, one should say, Hilbert's coordinates,
revealed itself to be a
genuine singularity of the manifold, defined in an invariant, local and
intrinsic way through the pure knowledge of the metric.
These facts are testified in any modern textbook by the
exhibition of the polinomial invariants built with the metric,
with the Riemann tensor and its covariant derivatives.
Therefore one might think that while Hilbert, thanks to
his error, stumbled over the right manifold, Schwarzschild's
conscious choice led him astray, due to the rudimentary status
in which the differential geometry of his times was still lying.
However, it will be noticed here that, despite the generally accepted opinion
reported above, Hilbert's manifold appears to be afflicted with two defects,
that are absent in Schwarzschild's manifold.

One of them was first taken care of by Synge, when he built from
Hilbert's manifold a clever geometric construction \cite{Synge50},
later mimicked by Kruskal and Szekeres in their maximal
extensions \cite{Kruskal60, Szekeres60},
in which the defect is no longer apparent. The shortcoming was
later explained by Rindler \cite{Rindler90, Rindler2001} to be at the
origin of the strange duplication that the maximal extensions
exhibit, with their bifurcate horizon and the unphysical prediction of the
necessary coexistence of both a future and a past singularity.

The defect is simply told: Hilbert's manifold intrinsically
disallows a consistent drawing of the time arrow;
only the change of topology induced by either the Synge
or by the Kruskal-Szekeres transformation with the inherent
redoubling allows one to get a manifold where the arrow
of time can be drawn without contradiction,
in keeping with Synge's postulates \cite{Synge50}.

A second defect of Hilbert's manifold is revealed \cite{AL2001,ALM2001}
with the contention that an invariant, local,
intrinsic singularity is found at Schwarz\-schild's two-surface,
provided that one does not limit the search, in this algebraically
special manifold, to the singularities exhibited by the invariants
build with the metric and with the Riemann tensor.
In Schwarzschild's manifold, and in the $r>2m$
part of Hilbert's manifold as well, through any event
one can draw a unique path of absolute rest, because
at each event the Killing equations
\begin{equation}\label{8}
\xi_{i;k}+\xi_{k;i}=0,
\end{equation}
and the condition of hypersurface orthogonality,
\begin{equation}\label{9}
\xi_{[i}\xi_{k,l]}=0,
\end{equation}
determine a unique timelike Killing vector $\xi_{i}$,
that therefore uniquely identifies the direction of
time. From $\xi_{i}$ one can define the four-velocity
\begin{equation}\nonumber
u^i\equiv\frac{\xi^i}{(\xi_i\xi^i)^{1/2}},
\end{equation}
the four-acceleration
\begin{equation}\nonumber
a^i\equiv\frac{\DD u^i}{\dd s}
\end{equation}
as absolute derivative of $u^i$ along its own direction,
and the norm of this four-acceleration
\begin{equation}\label{10}
\alpha=(-a_ia^i)^{1/2}.
\end{equation}
By using, say, Hilbert's manifold and coordinates,
$\alpha$ comes to read
\begin{equation}\label{11}
\alpha=\frac {m}{r^{3/2}(r-2m)^{1/2}}.
\end{equation}
Hence it diverges in the limit when $r\rightarrow 2m$.
Is not this divergence an invariant, local, intrinsic
singularity on the inner border of Schwarzschild manifold,
but, alas, in the interior of the Hilbert and of the
Kruskal manifolds?

\section{Appendix A: Frank's review of Schwarzschild's
``Massenpunkt'' paper}
For arbitrary gravitational fields, the Author deals with the
problem solved by {\it Einstein} for weak fields.
He looks for a solution of the field equation satisfying the
conditions that all the $g_{ik}$ be independent of $x_4$, that
$g_{14}=g_{24}=g_{34}=0$, that the solution be
spherically symmetric, and that the field vanish at infinity.
Then, in polar coordinates, the line element must have the form
$$
\dd s^2=F\dd t^2- (G+Hr^2)\dd r^2 - Gr^2(\dd \vartheta^2
+ \sin^2 \vartheta \dd \varphi^2),
$$
where $F, G, H$ are functions of $r$.
If one poses
$$ x_1=\frac{r^3}{3}, \quad x_2=- \cos \vartheta,
\quad x_3=\varphi, $$
it will be
$$ \dd s^2=f_4\dd x_4^2-f_1\dd x_1^2 - f_2\frac{\dd x_2^2}{1-x_2^2}
- f_3\dd x_3^2 (1-x_2^2),
$$
where
$$ f_2=f_3,
\quad f_4=F, \quad f_1=\frac{G}{r^4}+\frac{H}{r^2}, \quad f_2=Gr^2.
$$
Then, through integration of the field equations, it results
$$
f_1=\frac{1}{R^4} \frac{1}{1-\frac \alpha R}, \quad f_2=R^2,
\quad f_4=1-\frac \alpha R,
$$
where $R=\root 3\of{r^3+\alpha^3}$
and $\alpha$ is a constant, that depends on the mass of the point.
Therefore it is:
$$ \dd s^2= \left( 1-\frac \alpha R \right) \dd t^2 -
\frac{\dd R^2}{1- \frac \alpha R}-R^2 (\dd \vartheta^2
+ \sin^2 \vartheta \dd \varphi^2).
$$
For the equations of motion it results as first integral:
$$ \left( \frac{\dd x}{\dd \varphi} \right)^2 = \frac{1-h}{c^2}
+ \frac{h \alpha}{c^2} x-x^2+\alpha x^3,
$$
where $x=\frac 1R$, and $h$ is a constant of integration. If we substitute for $R$
its approximate value $r$, from this equation the one found by {\it Einstein}
is obtained, from which it results the motion of the
perihelion of Mercury. If by $n$ we mean the angular velocity of revolution,
according to the exact solution the third {\it Kepler}'s law
reads:
$$ n^2=\frac{\alpha}{2(r^3+\alpha^3)}.
$$
The proportionality between $n^2$ and $r^{-3}$ therefore does not hold exactly;
$n$ does not grow without limit for decreasing $r$,
but approaches itself to the maximal value $\frac{1}{\alpha \sqrt 2}$.

\section{Appendix B: Hilbert's derivation of the ``Schwarzschild'' metric}
\setcounter{equation}{41}
\noindent . . . . . . . . . . . . . . . . . . . . . . . . . . . . . . . .
 . . . . . . . . . . . . . . \par The integration
of the partial differential equations (36)
is possible also in another case, that for the first time has been dealt
with by Einstein\footnote{Perihelbewegung des Merkur, Sitzungsber. d. Akad.
zu Berlin. 1915, 831.} and by Schwarz\-schild\footnote{\"Uber das Gravitationsfeld
eines Massenpunktes, Sitzunsber. d. Akad. zu
Berlin. 1916, 189.}. In the following I provide for this case a
way of solution that does not make any hypothesis on the
gravitational potentials $g_{\mu\nu}$ at infinity, and that
moreover offers advantages also for my further investigations.
The hypotheses on the $g_{\mu\nu}$ are the following:
\begin{enumerate}
\item The interval is referred to a Gaussian coordinate system -
however $g_{44}$ will still be left arbitrary; {\it i.e.} it is\par
\smallskip
\centerline{$g_{14}=0, \ \ \ g_{24}=0, \ \ \ g_{34}=0.$}
\smallskip
\item The $g_{\mu\nu}$ are independent of the time coordinate $x_{4}$.
\item The gravitation $g_{\mu\nu}$ has central symmetry with
respect to the origin of the coordinates.
\end{enumerate}

According to Schwarzschild,
if one poses
\begin{eqnarray}\nonumber
w_{1}=r\cos\vartheta\\\nonumber
w_{2}=r\sin\vartheta \cos \varphi\\\nonumber
w_{3}=r\sin\vartheta \sin \varphi\\\nonumber
w_{4}=l
\end{eqnarray}
the most general interval corresponding to these hypotheses is represented
in spatial polar coordinates by the expression
\begin{equation} \label{42}
F(r)\dd r^{2}+G(r)(\dd \vartheta^{2}+\sin^{2}\vartheta \dd \varphi^{2})+H(r)\dd l^{2},
\end{equation}
where $F(r)$, $G(r)$, $H(r)$ are still arbitrary functions of $r$.
If we pose
\[
r^{\ast}=\sqrt{G(r)},
\]
we are equally authorised to interpret $r^{\ast}$,
$\vartheta$, $\varphi$ as spatial polar coordinates. If we
substitute in (\ref{42}) $r^{\ast}$ for $r$ and then drop the
symbol $\ast $, it results the expression
\begin{equation}
M(r)\dd r^{2}+r^{2}\dd \vartheta^{2}+r^{2}\sin^{2}\vartheta \dd \varphi^{2}+W(r)\dd l^{2},
\label{43}
\end{equation}
where $M(r)$, $W(r)$ mean the two essentially arbitrary functions of $r$.
The question is how the latter shall be determined in the most
general way, so that the differential equations (36) happen to be
satisfied.

To this end the known expressions $K_{\mu\nu}$, $K$, given in my
first communication, shall be calculated. The first step of this
task consists in writing the differential equations of the
geodesic line through variation of the integral
\[
\int {\left( M\left(\frac{\dd r}{\dd p}\right)^{2}+r^{2}\left(
\frac{\dd \vartheta}{\dd p}\right)^{2}+r^{2}\sin^{2}\vartheta\left(
\frac{\dd \varphi}{\dd p}\right)^{2}+W(\left(\frac{\dd l}{dp}\right)
^{2}\right) \dd p}.
\]
We get as Lagrange equations:
\begin{eqnarray}\nonumber
\frac{\dd^{2}r}{\dd p^{2}}+\frac{M^{\prime}}{2M}\left(\frac{\dd r}{\dd p}
\right)^{2}
-\frac{r}{M}\left[\left(\frac{\dd \vartheta}{\dd p}\right)^{2}
+\sin^{2}\vartheta\left(\frac{\dd \varphi}{\dd p}\right)^{2}\right]
-\frac{W^{\prime}}{2M}\left(\frac{\dd l}{\dd p}\right)^{2}=0,\\\nonumber
\frac{\dd^{2}\vartheta}{\dd p^{2}}+\frac{2}{r}\frac{\dd r}{\dd p}
\frac{\dd \vartheta}{\dd p}-\sin
\vartheta \cos\vartheta\left(\frac{\dd \varphi}{\dd p}\right)^{2}=0,\\\nonumber
\frac{\dd^{2}\varphi}{\dd p^{2}}+\frac{2}{r}\frac{\dd r}{\dd p}
\frac{\dd \varphi}{\dd p}+2\cot
\vartheta\frac{\dd \vartheta}{\dd p}\frac{\dd \varphi}{\dd p}=0,\\\nonumber
\frac{\dd^{2}l}{\dd p^{2}}+\frac{W^{\prime}}{W}\frac{\dd r}{\dd p}\frac{\dd l}{\dd p}=0;
\end{eqnarray}
here and in the following calculation the symbol $^{\prime}$ means
differentiation with respect to $r$. By comparison with the
general differential equations of the geodesic line:

\[
\frac{\dd^{2}w_{s}}{\dd p^{2}}+\sum_{\mu\nu}\left\{_{~s}^{\mu~\nu}\right\}
\frac{\dd w_{\mu}}{\dd p}\frac{\dd w_{\nu}}{\dd p}=0
\]
we infer for the bracket symbols $\left\{ _{~s}^{\mu~\nu}\right\}$
the following values (the vanishing ones are omitted):
\[
\left\{ _{~1}^{1~1}\right\} =\frac{1}{2}\frac{M^{\prime}}{M},\ \ \ \left\{
_{~1}^{2~2}\right\} =-\frac{r}{M},\ \ \ \left\{ _{~1}^{3~3}\right\}
=-\frac{r}{M}\sin^{2}\vartheta ,
\]
\[
\left\{ _{~1}^{4~4}\right\} =-\frac{1}{2}\frac{W^{\prime}}{M},\ \ \ \left\{
_{~2}^{1~2}\right\} =\frac{1}{r},\ \ \ \left\{ _{~2}^{3~3}\right\} =-\sin\vartheta
\cos\vartheta ,
\]
\[
\left\{ _{~3}^{1~3}\right\} =\frac{1}{r},\ \ \ \left\{ _{~3}^{2~3}\right\}
=\cot\vartheta, \ \ \ \left\{ _{~4}^{1~4}\right\}
=\frac{1}{2}\frac{W^{\prime}}{W}.
\]
With them we form:
\[
K_{11}=\frac{\partial}{\partial r}\left(\left\{ _{~1}^{1~1}\right\} +\left\{
_{~2}^{1~2}\right\} +\left\{ _{~3}^{1~3}\right\} +\left\{ _{~4}^{1~4}\right\}
\right) -\frac{\partial}{\partial r}\left\{ _{~1}^{1~1}\right\}
\]
\[
+\left\{ _{~1}^{1~1}\right\}\left\{ _{~1}^{1~1}\right\} +\left\{
_{~2}^{1~2}\right\}\left\{ _{~2}^{2~1}\right\} +\left\{ _{~3}^{1~3}\right\}
\left\{ _{~3}^{3~1}\right\} +\left\{ _{~4}^{1~4}\right\}\left\{
_{~4}^{4~1}\right\}
\]
\[
-\left\{ _{~1}^{1~1}\right\}\left(\left\{ _{~1}^{1~1}\right\}
+\left\{ _{~2}^{1~2}\right\} +\left\{ _{~3}^{1~3}\right\} +\left\{
_{~4}^{1~4}\right\} \right)
\]
\[
=\frac{1}{2}\frac{W^{\prime \prime}}{W}+\frac{1}{4}\frac{W^{\prime 2}}{W^{2}}
-\frac{M^{\prime}}{rM}-\frac{1}{4}\frac{M^{\prime}W^{\prime}}{MW}
\]
\smallskip
\[
K_{22}=\frac{\partial}{\partial\vartheta}\left\{ _{~3}^{2~3}\right\}
-\frac{\partial}{\partial r}\left\{ _{~1}^{2~2}\right\}
\]
\[
+\left\{ _{~2}^{2~1}\right\}\left\{ _{~1}^{2~2}\right\} +\left\{
_{~1}^{2~2}\right\}\left\{ _{~2}^{1~2}\right\} +\left\{ _{~3}^{2~3}\right\}
\left\{ _{~3}^{3~2}\right\}
\]
\[
-\left\{ _{~1}^{2~2}\right\}\left(\left\{
_{~1}^{1~1}\right\}+\left\{ _{~2}^{1~2}\right\} +\left\{
_{~3}^{1~3}\right\} +\left\{ _{~4}^{1~4}\right\} \right)
\]
\[
=-1-\frac{1}{2}\frac{rM^{\prime}}{M^{2}}+\frac{1}{M}+\frac{1}{2}\frac{%
rW^{\prime}}{MW}
\]
\smallskip
\[
K_{33}=-\frac{\partial}{\partial r}\left\{ _{~1}^{3~3}\right\} -\frac{%
\partial}{\partial\vartheta}\left\{ _{~2}^{3~3}\right\}
\]
\[
+\left\{ _{~3}^{3~1}\right\}\left\{ _{~1}^{3~3}\right\} +\left\{
_{~3}^{3~2}\right\}\left\{ _{~2}^{3~3}\right\} +\left\{ _{~1}^{3~3}\right\}
\left\{ _{~3}^{1~3}\right\} +\left\{ _{~2}^{3~3}\right\}\left\{
_{~3}^{2~3}\right\}
\]
\[
-\left\{ _{~1}^{3~3}\right\}\left(\left\{ _{~1}^{1~1}\right\}
+\left\{ _{~2}^{1~2}\right\} +\left\{ _{~3}^{1~3}\right\} +\left\{
_{~4}^{1~4}\right\} \right)-\left\{ _{~2}^{3~3}\right\}\left\{
_{~3}^{2~3}\right\}
\]
\[
=\sin^{2}\vartheta\left( -1-\frac{1}{2}\frac{rM^{\prime}}{M^{2}}+\frac{1}{M}+\frac{%
1}{2}\frac{rW^{\prime}}{MW}\right)
\]
\smallskip
\[
K_{44}=-\frac{\partial}{\partial r}\left\{ _{~1}^{4~4}\right\} +\left\{
_{~4}^{4~1}\right\}\left\{ _{~1}^{4~4}\right\} +\left\{ _{~1}^{4~4}\right\}
\left\{ _{~4}^{4~1}\right\}
\]
\[
-\left\{ _{~1}^{4~4}\right\}\left(\left\{ _{~1}^{1~1}\right\}
+\left\{ _{~2}^{1~2}\right\} +\left\{ _{~3}^{1~3}\right\} +\left\{
_{~4}^{1~4}\right\} \right)
\]
\[
=\frac{1}{2}\frac{W^{\prime \prime}}{M}-\frac{1}{4}\frac{M^{\prime
}W^{\prime}}{M^{2}}-\frac{1}{4}\frac{W^{\prime 2}}{MW}+\frac{W^{\prime}}{rM%
}
\]
\smallskip
\[
K=\sum_{s}g^{ss}K_{ss}=\frac{W^{\prime \prime}}{MW}-\frac{1}{2}%
\frac{W^{\prime 2}}{MW^{2}}-2\frac{M^{\prime}}{rM^{2}}-\frac{1}{2}\frac{%
M^{\prime}W^{\prime}}{M^{2}W}-\frac{2}{r^{2}}+\frac{2}{r^{2}M}+2\frac{%
W^{\prime}}{rMW}.
\]
Since
\[
\sqrt{g}=\sqrt{MW}r^{2}\sin\vartheta
\]
it is found
\[
K\sqrt{g}=\left\{\left(\frac{r^{2}W^{\prime}}{\sqrt{MW}}\right)^{\prime
}-2\frac{rM^{\prime}\sqrt{W}}{M^{\frac{3}{2}}}-2\sqrt{MW}+2\sqrt{\frac{W}{M}%
}\right\} \sin\vartheta \
\]
and, if we set
\[
M=\frac{r}{r-m},\ \ \ \ W=w^{2}\frac{r-m}{r},\
\]
where henceforth $m$ and $w$ become the unknown functions of $r$,
we eventually obtain
\[
K\sqrt{g}=\left\{\left(\frac{r^{2}W^{\prime}}{\sqrt{MW}}\right)^{\prime}
-2wm^{\prime}\right\} \sin\vartheta .\
\]
Therefore the variation of the quadruple integral
\[
\int\int\int\int K\sqrt{g}\dd r\dd \vartheta \dd \varphi \dd l\
\]
is equivalent to the variation of the single integral
\[
\int wm^{\prime}\dd r\
\]
and leads to the Lagrange equations
\begin{eqnarray}\label{44}
m^{\prime}=0,\\\nonumber
w^{\prime}=0.
\end{eqnarray}
One easily satisfies oneself that these equations effectively entail the
vanishing of all the $K_{\mu\nu}$; they represent therefore
essentially the most general solution of the equations (36) under
the hypotheses (1), (2), (3) previously made.
 If we take as integrals of (\ref{44}) $m=\alpha$,
where $\alpha$ is a constant, and $w=1$ (a choice that evidently does
not entail any essential restriction)
from (\ref{43}) with $l=it$ it results the looked for interval
in the form first found by Schwarzschild

\begin{equation}
G(\dd r,\dd \vartheta ,\dd \varphi ,\dd l)
=\frac{r}{r-\alpha}\dd r^{2}+r^{2}\dd \vartheta
^{2}+r^{2}\sin^{2}\vartheta \dd \varphi^{2}-\frac{r-\alpha}{r}\dd t^{2}.
\label{45}
\end{equation}
The singularity of this interval for $r=0$ vanishes only when it is
assumed $\alpha=0$, {\it i.e.}: {\gr under the hypotheses (1), (2), (3)
the interval of the pseudo-Euclidean geometry is the only regular interval
that corresponds to a world without electricity.}

For $\alpha\neq 0$,  $r=0$ and, with positive values of $\alpha$, also
$r=\alpha$ happen to be such points that in them the interval is not regular.
I call an interval or a gravitational field $g_{\mu\nu}$ {\it regular}
in a point if, through an invertible one to one transformation, it is possible
to introduce a coordinate system such that for it the
corresponding functions $g_{\mu\nu}^{\prime}$ are regular in that
point, {\it i.e.} in it and in its neighbourhood they are
continuous and differentiable at will, and have a determinant
$g^{\prime}$ different from zero.

Although in my opinion only regular solutions of the fundamental
equations of physics immediately represent the reality,
nevertheless just the solutions with non regular points are an
important mathematical tool for approximating characteristic
regular solutions - and in this sense, according to the procedure
of Einstein and Schwarzschild, the interval (\ref{45}), not regular
for $r=0$ and for $r=\alpha$, must be considered as expression
of the gravitation of a mass distributed with central symmetry
in the surroundings of the origin\footnote{Transforming to the origin
the position $r=\alpha$, like Schwarzschild did, is in my opinion not
advisable; moreover Schwarz\-schild's transformation is not the
simplest one, that reaches this scope.}. . . . . . . . . . .
. . . . . . . . . . . . . . . . . . . . . .
 . . . . . . .

\end{document}